\documentclass[conference]{IEEEtran}
\IEEEoverridecommandlockouts

\usepackage{cite}
\usepackage{amsmath,amssymb,amsfonts}
\usepackage{algorithmic}
\usepackage{mathtools, nccmath}
\usepackage{graphicx}
\usepackage{color,soul}
\usepackage{textcomp}
\usepackage{xcolor}
\usepackage{gensymb}
\usepackage{nidanfloat}
\usepackage{subcaption}

\begin{document}

\title{Addressing Resiliency of In-Memory Floating Point Computation}

\author{\IEEEauthorblockN{Sina Sayyah Ensan, Swaroop Ghosh, Seyedhamidreza Motaman, and Derek Weast}
\IEEEauthorblockA{\textit{School of Electrical Engineering and Computer Science, Pennsylvania State University, University Park, PA 16802 USA} \\
(sxs2541, szg212, sxm884, dqw5347)@psu.edu}
}

\maketitle

\begin{abstract}

In-memory computing (IMC) can eliminate the data movement between processor and memory which is a barrier to the energy-efficiency and performance in Von-Neumann computing. Resistive RAM (RRAM) is one of the promising devices for IMC applications (e.g. integer and Floating Point (FP) operations and random logic implementation) due to low power consumption, fast operation and small footprint in crossbar architecture. In this paper, we propose FAME, a pipelined FP arithmetic ($adder$/$subtractor$) using RRAM crossbar based IMC. A novel shift circuitry is proposed to lower the shift overhead during FP operations. Since $96\%$ of the RRAMs used in our architecture are in High Resistance State (HRS), we propose two approaches namely Shift-At-The-Output (SATO) and Force To $V_{DD}$  (FTV) (ground (FTG)) to mitigate Stuck-at-1 (SA1) failures. In both techniques, the fault-free RRAMs are exploited to perform the computation by using an extra clock cycle. Although performance degrades by 50\%, SATO can handle 50\% of the faults whereas FTV can handle 99\% of the faults in the RRAM-based compute array at low power and area overhead. Simulation results show that the proposed single precision FP adder consumes $335$ pJ and $322$ pJ for $NAND$-$NAND$ and $NOR$-$NOR$ based implementations, respectively. The area overheads of SATO and FTV are 28.5\% and 9.5\%, respectively.

\end{abstract}


\section{introduction}

In the big data era, conventional CMOS-based Von-Neumann architecture platforms are unable to face real-time data processing requirements \cite{COFT}. Memory and computing elements are decoupled from each other in Von-Neumann architecture \cite{imaging} which apply frequent communication between memory and computing cores \cite{beyond}. The compute energy has been scaled asymmetrically compared to data transport energy with transistor scaling. Data movement in modern computing systems dominates energy-efficiency and performance \cite{roy}. 

In Memory Computing (IMC) is one of the promising compute models to fully or partially eliminate the need to transport data between processors and memory. The main concept of IMC is to infuse compute capability into the memory cells \cite{imani}. IMC is achievable by using emerging Non-Volatile Memories (NVM)  e.g., RRAM, Spin Transfer Torque (STT) RAM and Phase Change Memory (PCM) \cite{imani}, \cite{machine6t}, \cite{hamid-iccd}, \cite{deepsram}. 
Near memory processing \cite{iram} and logic-in-memory, which employ NVMs in the logic space \cite{fefet}, \cite{mtj} to preserve states between powering sequence have been proposed in the literature. However, they cannot solve the problem of separation between logic and memory.

IMC modifies memory cells and/or peripheral circuits/access mechanisms to infuse compute capability into memory cells. IMC can solve specific tasks such as, dot-products for recognition \cite{deepsram}, search \cite{ftcam} and classification \cite{machine6t}. It also supports a wide range of logic and arithmetic operations \cite{fefet}, \cite{mpim}, \cite{drambit}, \cite{IJCNN}. NVM-based IMC using STTRAM \cite{bitstt}, RRAM \cite{magicjo}, Ferroelectric FET (FeFET) and Phase-Change Memory \cite{pinatubo} are becoming popular.

Due to immature fabrication technology limitations, manufacturing yield is still a serious concern for NVMs such as, RRAM crossbar. Faults in RRAM crossbar arrays are categorized into hard and soft faults \cite{COFT}. Previous studies have been predominantly focused on soft faults \cite{ice} whereas few attempts are made to recover crossbar arrays from hard faults. The soft faults (e.g., read disturb) can be recovered by calibrating the resistance \cite{ice} \cite{explore}. However, hard faults are recovered through mapping algorithms (i.e., by assigning inputs of faulty RRAMs to the redundant rows or columns) \cite{COFT}, \cite{stuck}, \cite{handling}. 

Stuck-at fault is defined as a situation when the RRAM is permanently stuck at High Resistance State (HRS) or Low Resistance State (LRS). It has been reported \cite{failure} that only 63\% of $HfO_2-$ based RRAM devices for 4Mb crossbar array are fault-free and about 10\% of RRAM devices contain stuck-at faults. Retention failure which is similar to the resistive switching due to the generation or recovery of oxygen vacancy is another type of hard faults in RRAMs. In the proposed IMC architecture, only 4\% of the RRAMs are in LRS and the other 96\% are in HRS. Therefore in this paper, we focused on the HRS retention failure and stuck-at-1 (i.e., stuck-at HRS) faults. If the yield of a single RRAM device is 99\%, there is only $10^{-9}$ probability for a column of 64*32 array to be fault free. The stuck-at failures and HRS to LRS switching \cite{detection} can be fixed by employing few redundant rows/columns when RRAM array is considered working as a memory. However, the whole array is needed for IMC application. Consequently, computations will fail due to errors in the absence of fault tolerance schemes.

We have considered Floating Point (FP) operations to evaluate the proposed resilience techniques. This is motivated by the fact that emerging applications e.g., mission-critical systems like autonomous cars require huge amount of data processing in real-time at low-power (to make timely decisions). The autonomous cars make complex decisions in a tight deadline using algorithms e.g., Kalman filters for data fusion, ray tracing for path planning and, edge detection and deep neural networks for classification. Most of these algorithms require FP vector operations involving transpose, inverse, addition/multiplication. Therefore, the capability to perform these tasks, quickly and accurately can be of utmost importance to enable the safe and energy-efficient autonomous systems. Conventionally, FP architectures are implemented as full custom VLSI or in FPGA. Although fast and power efficient, these custom designs impose cost and complexity. In this paper, we propose FAME (Single Precision Floating Point Arithmetic using In-Memory Computing) implemented on crossbar RRAM. We employ a modified version (Section \ref{baseline}) of Dynamic Computing In Memory (DCIM) \cite{hamid-iccd} based architecture as our baseline compute substrate for FAME. Additionally, two approaches namely, Shift-At-The-Output (SATO) and Force To $V_{DD}(GND)$ (FTV(G)) are proposed to enable in-memory computing in presence of HRS to LRS retention failures. We focus on this failure mechanism due to two reasons: (i) HRS to LRS switching is more common in RRAM\cite{gao2011modeling}; (ii) majority of the RRAMs (96\%) are in HRS for both $NAND$-$NAND$ and $NOR$-$NOR$ arrays. Carry Select Adder (CSA) based on DCIM implementation is used for the demonstration. We add extra peripheral circuits on each array to implement the proposed techniques. 

In particular, we make the following contributions in this paper:
\begin{enumerate}
\item Alternative low-overhead realization of DCIM for FP computation;
\item In-memory shift circuit embedded in the peripherals e.g., sense amplifier (SA); 
\item Enabling pipeline architecture using the latch embedded in the SA;
\item Propose fault mitigation approaches such as, SATO and FTV/FTG for DCIM architecture;
\item Conduct PV analysis of the RRAM array to check the integrity of SATO and FTV/FTG. 
 \end{enumerate}


Rest of the paper is organized as follows. Section \ref{related-work-and-background} introduces related work on IMC. Section \ref{FAME-Circuit-and-Architecture} explains the proposed FAME circuit and architecture. Section \ref{FAME-Simulation-Results} presents the simulation results of FAME and comparison with other IMC logic implementation. Section \ref{Proposed-Approaches-for-Stuck-at-fault-hanfdling} explains proposed approaches to overcome SA1 faults in IMC architecture. Section \ref{Fault-Simulation-Results} presents the proposed fault tolerance approaches and simulation results. Section \ref{Conclusion} draws the conclusion.

\begin{figure}
	\centering
	\includegraphics[width= \linewidth]{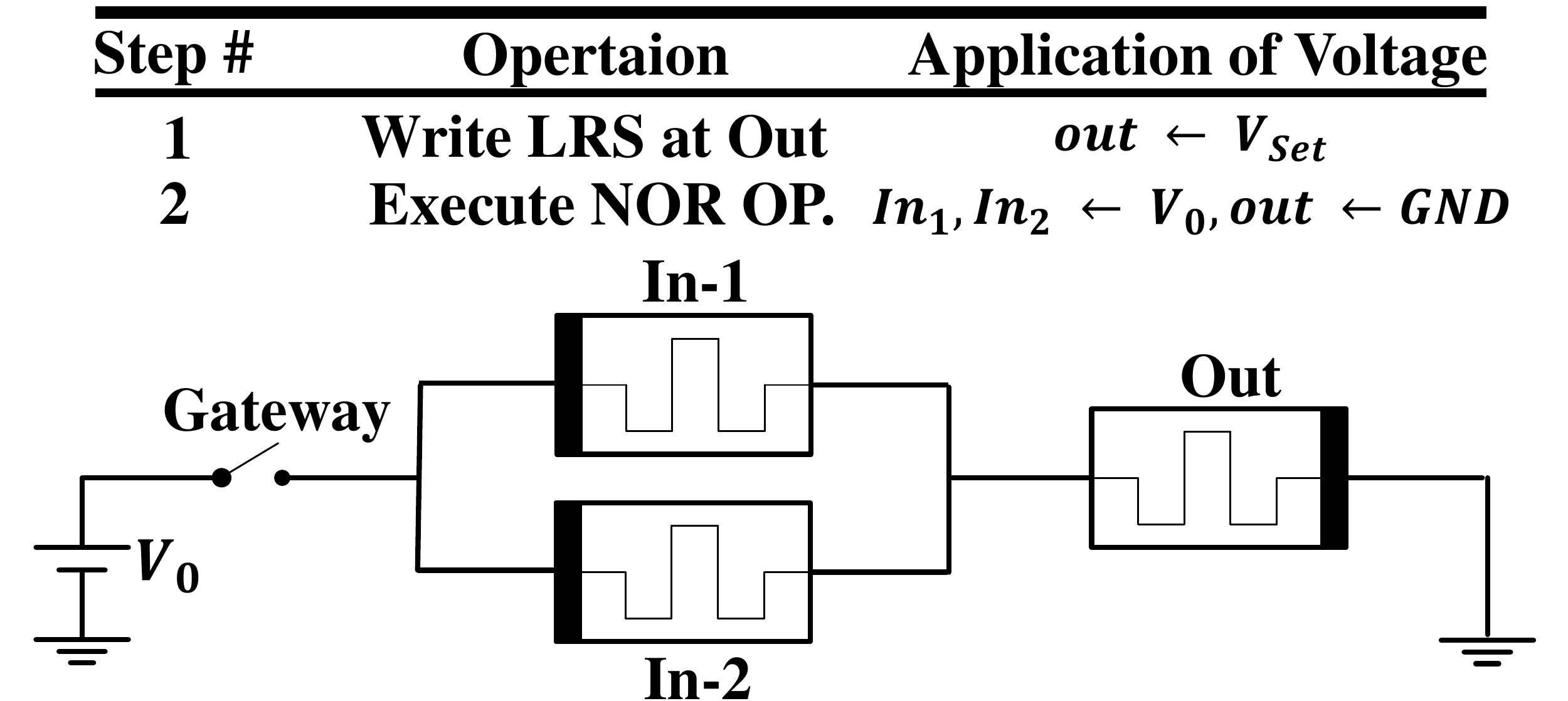}
	\caption{MAGIC $NOR$}
	\label{magic}
\end{figure}

\section{Related Work and Background} {\label{related-work-and-background}}

\subsection{Memristor Aided Logic (MAGIC)}
MAGIC \cite{magic} (shown in Fig. \ref{magic}) is an IMC architecture in which logic state of the gates are represented by the memristor (RRAM in this paper) resistance where high (low) resistance is considered as logic `1' (`0'). The inputs to a MAGIC gate are the logic states stored in the input memristors and the output is the final state of the output memristor. MAGIC executes operations in two steps: 1) setting the output memristor to a known logic state (e.g., for $NOR$ operation the output is in LRS); 2) applying a known voltage ($V_0$) to the input memristors which causes current flow through the input and output memristors. The output memristor's state changes if the current passing through it is higher than the set/reset current. MAGIC is capable of implementing Boolean functions such as, $NAND$, $NOR$, $AND$, $OR$ and $NOT$.

\subsection{Dynamic Computing In Memory (DCIM)}
\begin{figure*}
        \centering 
        \begin{subfigure}{0.63\linewidth}
                \centering
                \includegraphics[width=0.99\linewidth]{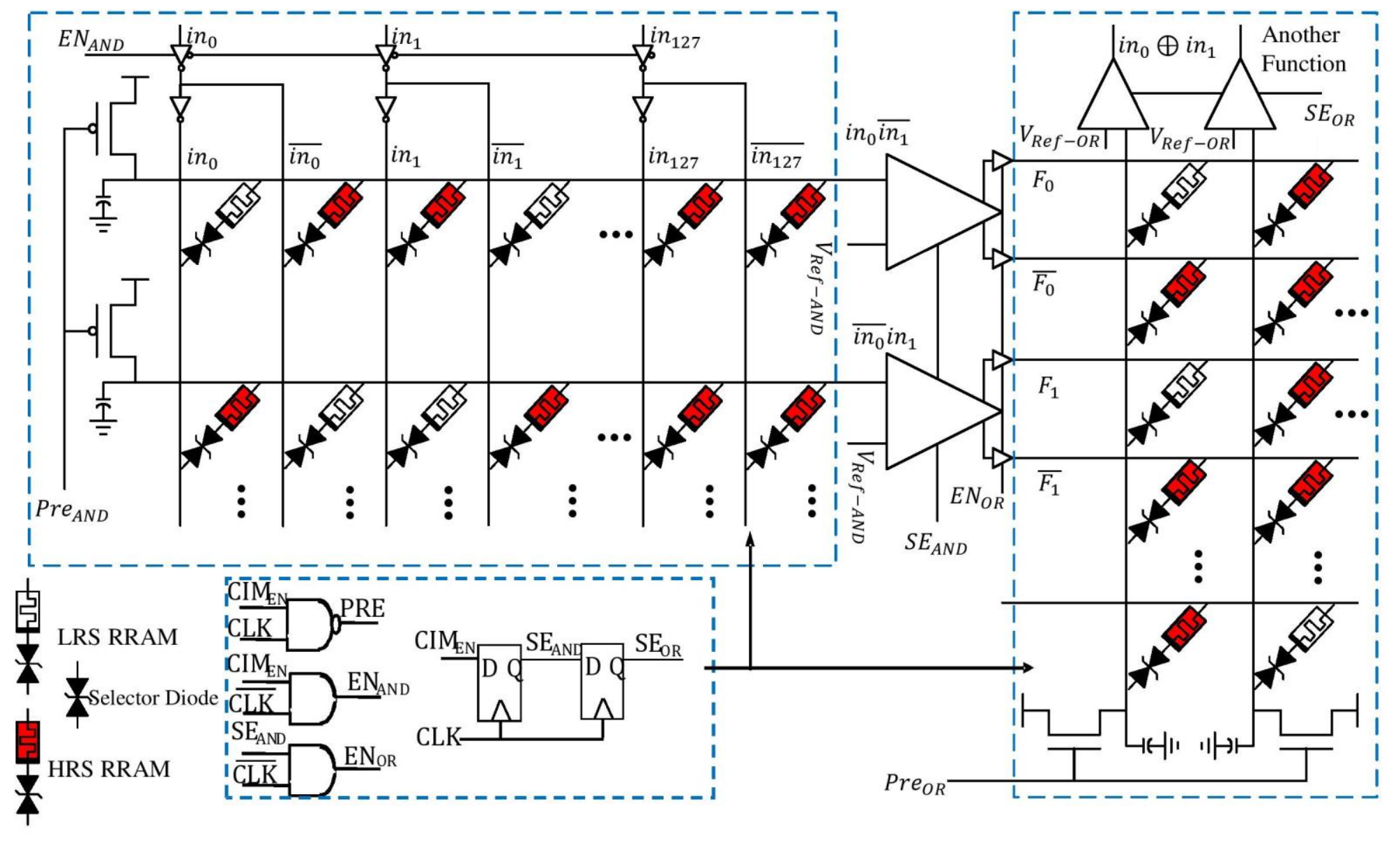}
                \caption{}
                \label{background-a}
        \end{subfigure}%
        \begin{subfigure}{0.37\linewidth}
                \centering
                \includegraphics[width=0.99\linewidth]{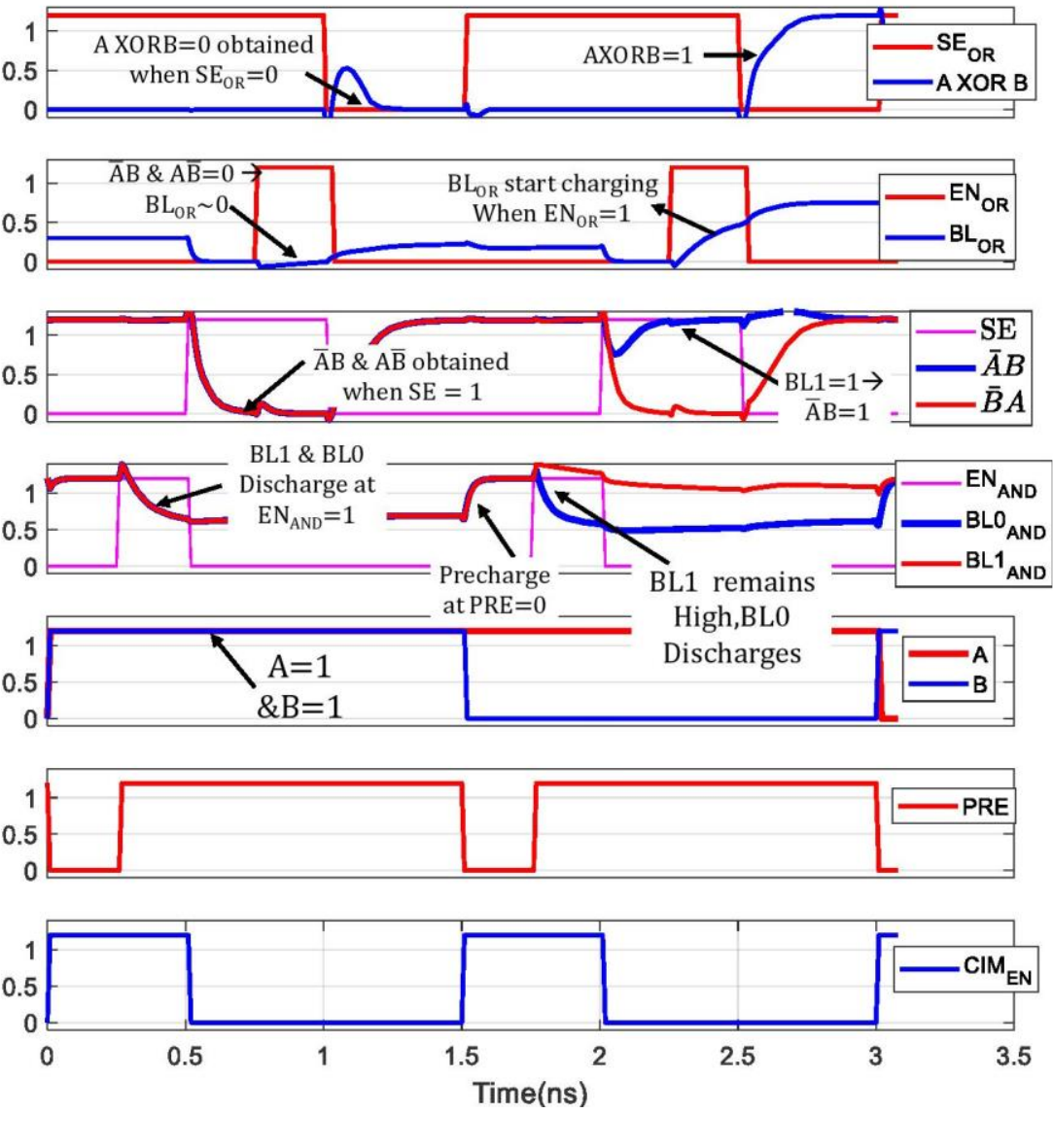}
                \caption{}
                \label{background-b}
        \end{subfigure}\\ 
        \caption{(a) XOR implementation using DCIM architecture in RRAM crossbar array; and, (b) timing diagram of logical XOR operation. }
        \label{background}
\end{figure*}

{\label{DCIM}}
DCIM \cite{hamid-iccd} is an RRAM crossbar based architecture, which each memory cell is composed of an RRAM device connected in series with a selector diode (Fig. \ref{background-a}. In-memory computation is accomplished by implementing the functions in the form of Sum-of-Product (SoP). Thus, both $AND$ and $OR$ operations are required to implement the logical functions. 

In DCIM, wordlines (WL) serve as the inputs and the bitlines (BL) serve as the outputs of the arrays. Separate pre-programmed $AND$ and $OR$ arrays are dedicated to implement the desired function. For instance, in order to implement $in_0.\overline{in_1}$, the bitcells connected to $in_0$ and $\overline{in_1}$ are programmed to LRS while the bitcells connected to $\overline{in_0}$ and $in_1$  are programmed to HRS (Fig. \ref{background-a}. All bitcells which are not part of $AND$ gate inputs are programmed to HRS (e.g., the bitcells connected to input $in_n$ and $\overline{in_n}$).

Fig. \ref{background} shows the implementation of $XOR$ function using DCIM. Initially, Pre signal is activated to pre-charge BLs of the $AND$ array. Next, inputs (${in_0}$ and $in_1$) are applied by asserting $EN_{AND}$. As shown in Fig. \ref{background-b}, both $BL_0$ and $BL_1$ drop below the reference voltage ($V_{Ref-AND}$) when $in_0=in_1=1$. As a result, SA output which determines the results of $in_0.\overline{in_1}$ and $\overline{in_0}.in_1$ functions are pulled down to `0' at the edge of $SE_{AND}$. Next, $AND$ array SA outputs are provided as inputs to the $OR$ array. Since inputs of the $OR$ array are `0', the BL ($BL0_{OR}$) remains discharged which results in $in_0\oplus{in_1}=0$. If $in_0=0, in_1=1$ ($in_0.\overline{in_1}=0$ and $\overline{in_0}.in_1=1$), $BL_0$ discharges while $BL_1$ remains pre-charged. Therefore, $BL0_{OR}$ starts charging at the edge of $EN_{OR}$. Finally, the voltage of $BL0_{OR}$ is compared against $V_{Ref-OR}$ at the edge of $SE_{OR}$ which produces `1' at the output of SA.

\subsection{FP Addition/Subtraction}

\begin{figure}
	\centering
	\includegraphics[width= \linewidth]{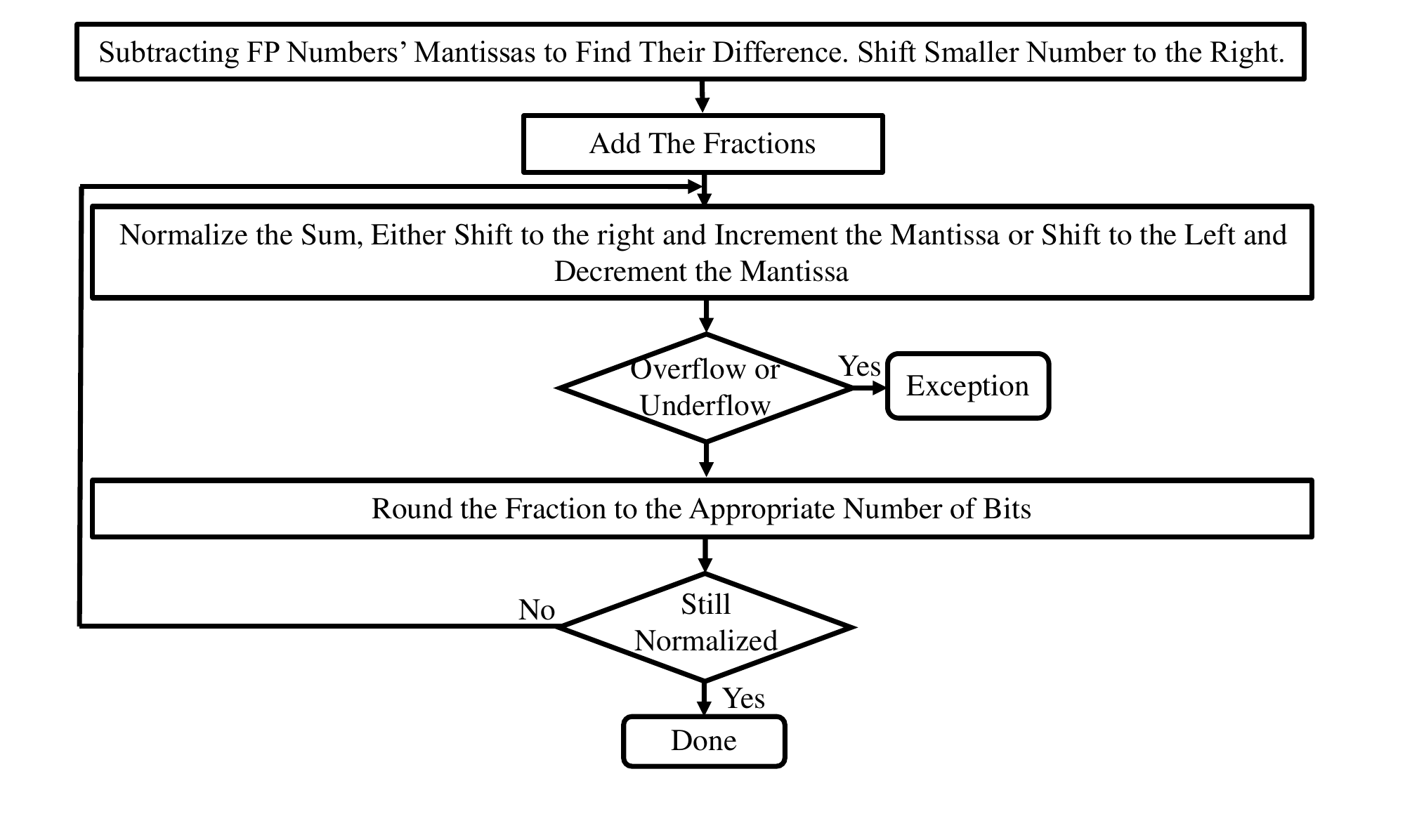}
	\caption{IEEE 754 Standard FP addition/subtraction Flowchart \cite{hennessy}}
	\label{flowchart}
\end{figure}

In IEEE 754 standard, a single precision FP number is represented by 1 Sign bit, 8 Exponent bits, and 23 Fraction bits. A negative (positive) number is represented with a sign bit equal to `1'(`0'). In order to demonstrate negative exponents, IEEE 754 uses a bias of 127 for single precision (e.g., -1 is represented by -1+127=126). The general representation of a FP number is given by:
  \begin{equation}
 \label{pipeline}
(-1)^{Sign} * (1+Fraction)*2^{(Exponent-Bias)} 
\end{equation} The flowchart for FP addition/subtraction as per IEEE 754 standard is shown in Fig. \ref{flowchart}. 

\section{FAME Simulation Results} {\label{FAME-Simulation-Results}}

The  simulations are carried out in 65nm PTM \cite{ptm} technology by employing ASU RRAM model\cite{asu} and bi-directional selector diode model \cite {selector}. Worst-case Sense Margin (SM), BL-delay, average delay, average power, and energy consumption (Table {\ref{results}}) are calculated to evaluate FAME architecture. Key parameters of devices for simulations are listed in Table {\ref{parameters}}. SM is obtained by performing 1000 point Monte Carlo simulations at various temperatures with parameters listed in Table \ref{variation} to mimic process variations. 

\begin{table}
    \centering
    \caption{Simulation parameters}
    \begin{tabular}{|c|c|}\hline
        Parameter & Value \\ \hline
        MOSFET Gate Length & 65 nm \\ \hline
        NMOS/PMOS Threshold Voltage & 423/-365 mV  \\ \hline
        BL Capacitance & 30 fF \\ \hline
        RRAM Gap Min/Max/Oxide Thickness & 0.1/1.7/5 nm \\ \hline
        \scriptsize{Atomic Energy for Vacancy Generation/Recombination} & 1.501/1.5 eV \\ \hline
        RRAM Write Latency & 25 ns\\ \hline
        RRAM HRS/LRS at 1.2V & 6.68 M/58.9 K $\Omega$ \\
        \hline\end{tabular}
        \label{parameters}
\end{table}

\begin{table}
    \centering
    \caption{Monte Carlo simulation parameters}
    \begin{tabular}{|c|c|c|c|}\hline
        Parameter & Real Value & Variation & STD. Deviation \\ \hline
        RRAM LRS Gap & 0.1 nm & 7\% & $3\sigma$\\ \hline
        RRAM HRS Gap & 1.7 nm & 7\% & $3\sigma$ \\ \hline
        MOS Oxide Thickness& 1.2 nm & 10\% & $3\sigma$ \\ \hline
        MOS Gate Length & 65nm & 10\% & $3\sigma$ \\ 
        \hline\end{tabular}
        \label{variation}
\end{table}

The worst-case SM is obtained under process variation $@25^oC$ for worst case compute array (i.e., fraction addition array). The BL-delay is the time when 100 mV SM is achieved. The proposed FP adder/subtractor implementation with both $NAND$-$NAND$ and $NOR$-$NOR$ architecture are compared against MAGIC and ASIC design. 

The write latency is obtained by performing 1000 points MC simulation. The worst-case write latency for low-to-high and high-to-low switching under process variation is 20ns. FAME achieves 828X, 3.2X and 3.7X improvement in latency, power and energy, respectively compared to MAGIC. The higher energy associated with MAGIC is attributed to the need to write into the RRAMs when an operation is done. Furthermore, compared to the power, energy consumption, and delay imposed by transferring data between main memory and processing units (e.g. CPU, GPU, and FPGA), FAME reduces power and energy consumption and delay by $98.8\%$, $93.7\%$, and $70.2\%$, respectively.

\begin{figure}
	\centering
	\includegraphics[width=250 pt]{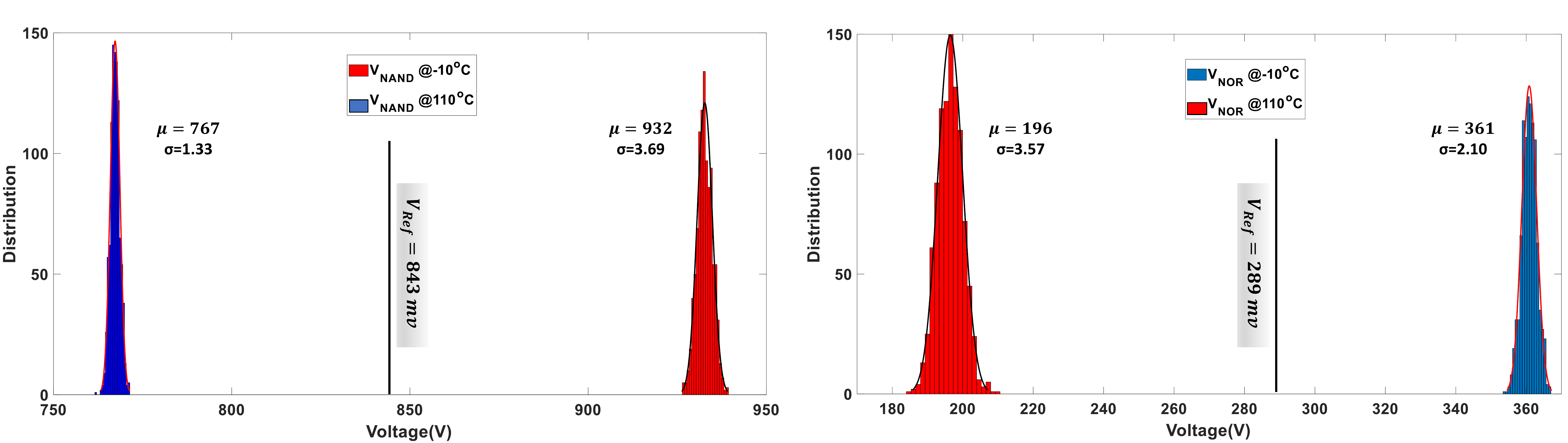}
	\caption{$SM$ distribution.}
	\label{sm-dist}
\end{figure}

A 1000 point MC simulations are performed at $-10\degree$C, $25\degree$C, and $90\degree$C at $1.2 V$ supply voltage to obtain mean of SM (Table \ref{results}). $V_{NAND0}$ (NAND array BL voltage when input is `0'), $V_{NAND1}$, $V_{NOR0}$, and $V_{NOR1}$ distributions at worst-case temperature are shown in Fig. \ref{sm-dist}. In order to achieve the read access pass yield ($RAPY$) \cite{rapy}\cite{hamid-iccd} we have performed SA offset voltage analysis. The SA offset voltage can be modeled by a Gaussian distribution with $\sigma=16mV$ and $\mu=8mV$. To obtain $RAPY$ we assume that $V_{Ref}$ is produced by a voltage regulator with negligible variation ($5mV$). We assigned  $V_{Ref}$ in such a way to maximize $RAPY$. Based on the Monte-Carlo simulation, the RAPY of $NAND$ and $NOR$ operations are found to be $4.6\sigma$ and $4.5\sigma$ respectively.

\begin{table}
    \centering
    \caption{Simulation results}
    \begin{tabular}{|c|c|c|c|c|}\hline
         {Characteristics} & {$NAND$} & {$NOR$} & {MAGIC} & CPU \cite{stanford} \\ \hline
        
         {BL Delay (ns)} & {1.42} & {1.23} & {N/A} & {N/A}\\ \hline
        
         {SA Sense Delay (ps)} & {24.52} & {69.1} & {N/A} & {N/A}\\ \hline
        
         {Average Delay} & {25ns} & {23ns} & {20us} & {84ns}\\ \hline
        
         {Exp. Subt. Pow. (uW)}& {443.31} & {448} & {2808.92} & N/A \\ \hline
        
         {Fr. Add. Pow. (uW)}& {1068.52} & {1123.19} & {2142.84} & N/A \\ \hline
        
         {Shift Pow. (uW)}& {443.24} & {452.93} & {982.31} & N/A \\ \hline
        
         {Avg. Power (mW)}& {0.7} & {0.71} & {2.3} & 61 \\ \hline
        
         {Energy (nJ)}& {0.33} & {0.32} & {1.2} & 5.1\\
        \hline\end{tabular}
        \label{results}
\end{table}

\begin{table}
    \centering
    \caption{SM in different temperatures}
    \begin{tabular}{|c|c|c|c|}\hline
        SM (mv) / Temp & $-10\degree$ & $25\degree$ & $90\degree$ \\ \hline
        $NAND$ & 94.5499 & 91.30245 & 79.26 \\ \hline
        $NOR$ & 105.6009 & 104.3965 & 99.7171 \\ \hline
        \end{tabular}
        \label{temperature-results}
\end{table}

\begin{table}
    \centering
    \caption{FAME area}

    \begin{tabular}{|c|c|c|}\hline
        Block & Array Size & \# of Arrays \\ \hline
        Exponent Subtraction 1st $NAND$ & 32*32 & 1\\ \hline
        Exponent Subtraction 2nd $NAND$ & 32*64 & 1 \\ \hline
        Right Shift & 8*16 & 1 \\ \hline
        Fraction Addition 1st $NAND$ & 64*64 & 2 \\ \hline
        Fraction Addition 2nd $NAND$ & 64*64 & 2 \\ \hline
        Left Shift & 32*64 & 1 \\ \hline
        Exponent Inc/Dec 1st $NAND$ & 32*32 & 1 \\ \hline
        Exponent Inc/Dec 1st $NAND$ & 32*64 & 1 \\ \hline

        \end{tabular}
        \label{FAME-Area}
\end{table}

\section{Resilience to Stuck-at fault} {\label{Proposed-Approaches-for-Stuck-at-fault-hanfdling}}

In this section, we describe SATO and FTV, two fault mitigation techniques proposed for DCIM architecture. In the following we use, (i) faulty BL to denote each BL with an undesired stuck-at-1 (SA1) RRAM; (ii) faulty WL (BL) to denote each WL (BL) with an undesired SA1 RRAM.

\begin{figure}
        \centering
        \begin{subfigure}[b]{0.5\textwidth}
                \centering
                \includegraphics[width=0.9\linewidth]{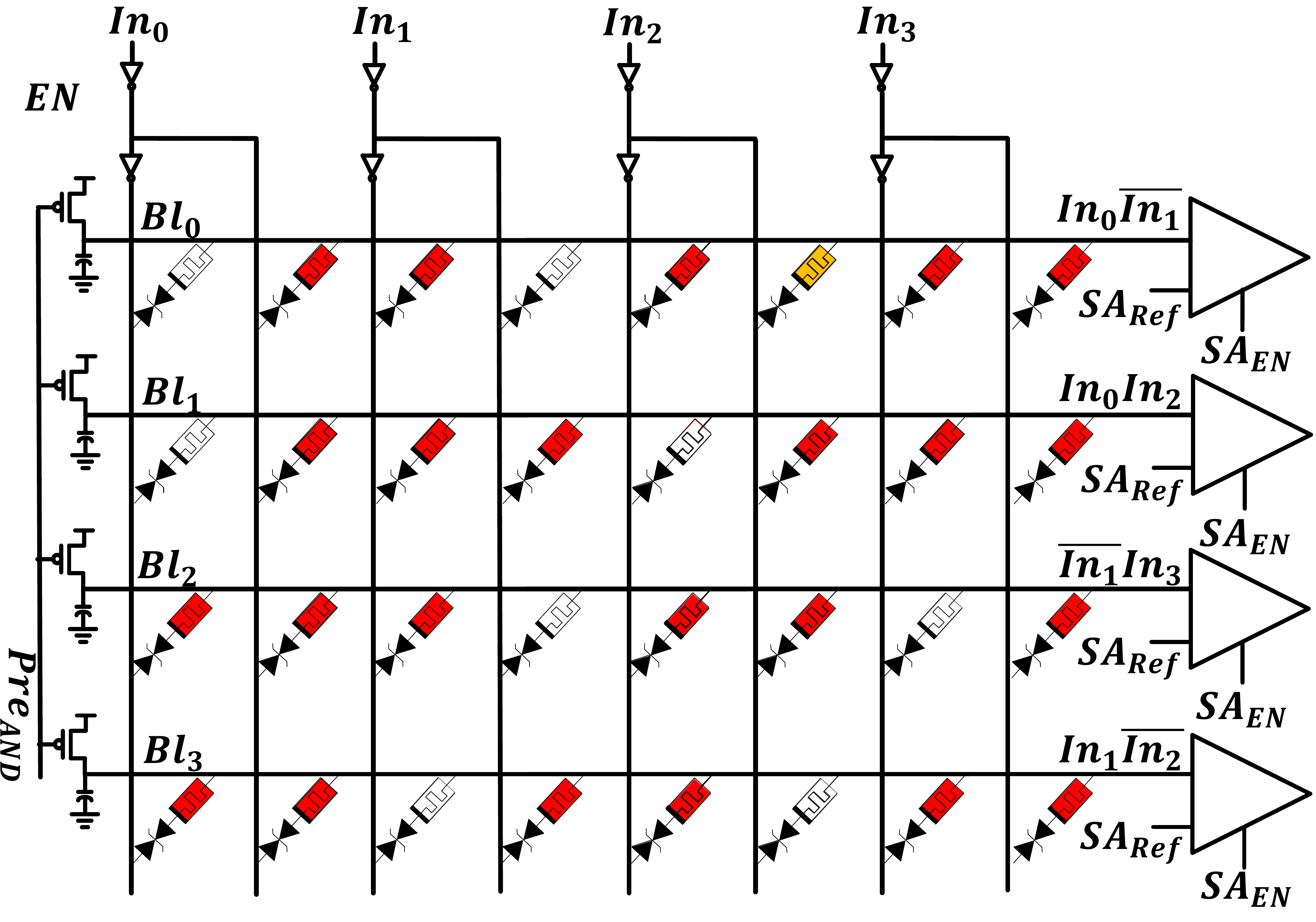}
                \caption{}
                \label{time-hl-a}
        \end{subfigure}\\
        \begin{subfigure}[b]{0.5\textwidth}
                \centering
                \includegraphics[width=0.99\linewidth]{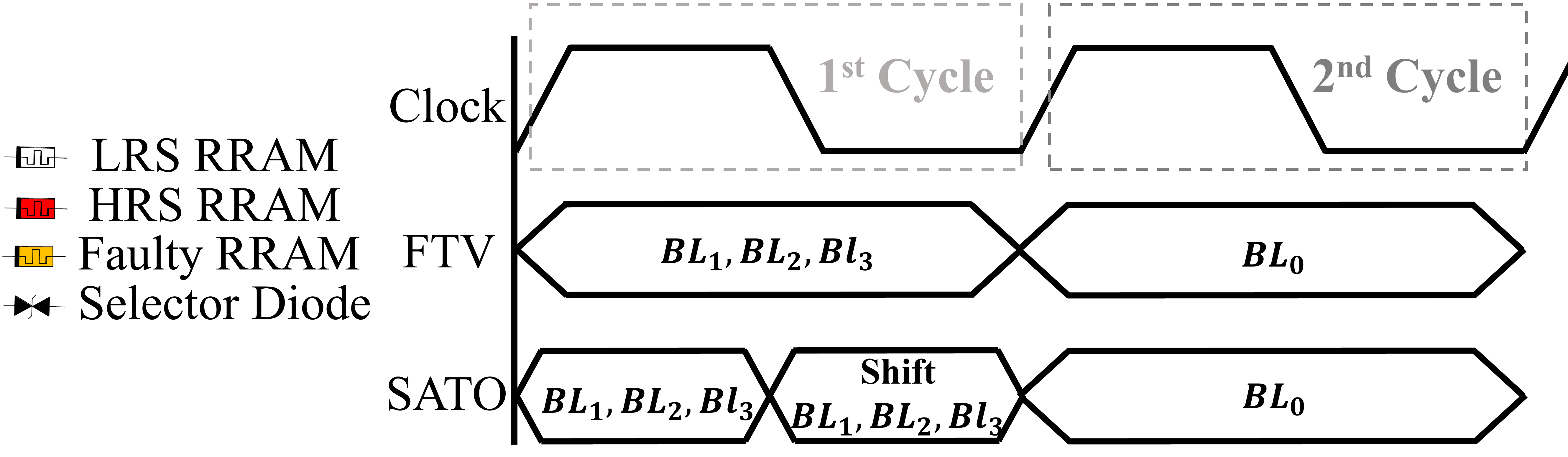}
                \caption{}
                \label{time-hl-b}
        \end{subfigure}%
        \caption{(a) 4*4 RRAM crossbar array, (b) FTV and SATO timing.}
        \label{time-hl}
\end{figure}


Computations are performed in two cycles when the proposed fault mitigation techniques are applied (Fig. \ref{time-hl}).
The computations of fault-free BLs ($BL_1$, $BL_2$ and $BL_3$ in Fig. {\ref{time-hl}} (a)) are performed in the first cycle and the computations of the faulty BLs ($BL_0$ in Fig. {\ref{time-hl}} (a)) are performed in the second cycle. In FTV, the WLs corresponding to faulty RRAMs ($\overline{In_2}$ in Fig. {\ref{time-hl}} (a)) for $NAND$ ($AND$) array are forced to $V_{DD}$ to mask faulty bits. In a dual Force-to-Ground (FTG) technique, the faulty BLs are forced to 0V for NOR (OR) arrays. FTV/FTG tolerates 99\% of stuck-at faults (SAF) while reducing power consumption of the array. In SATO approach, operations of fault-free BLs are executed in the first cycle and then the outputs are shifted in the SAs. Then, the operations of faulty BLs are computed using fault-free BLs (operation of $BL_0$ is done in $BL_1$). SATO covers 50\% of SAFs without affecting power consumption. The high level timings of FTV and SATO are illustrated in Fig. {\ref{time-hl}} (b) and (c), respectively.

\subsection{Shifting-At-The-Output (SATO)}

As described before, in this technique the normal operation for fault-free BLs are performed in the first cycle and computation of faulty BLs are performed in the second cycle. 
SATO does not use faulty BLs for performing an operation and executes all the operations on the fault-free BLs. SATO shifts the data stored in SAs' latch of fault-free BLs to prevent overwriting. When computation of first cycle is completed, the data are shifted in SAs (three shifts are needed if an adder/subtractor is implemented). As shown in Fig. \ref{SATO}, inputs of the WLs should get shifted too, so computation is performed using fault-free BLs. Peripherals of SATO incurs 29.5\% area overhead.

\subsubsection{Non-fixable Faults}
\begin{figure}
	\centering
	\includegraphics[width= \linewidth]{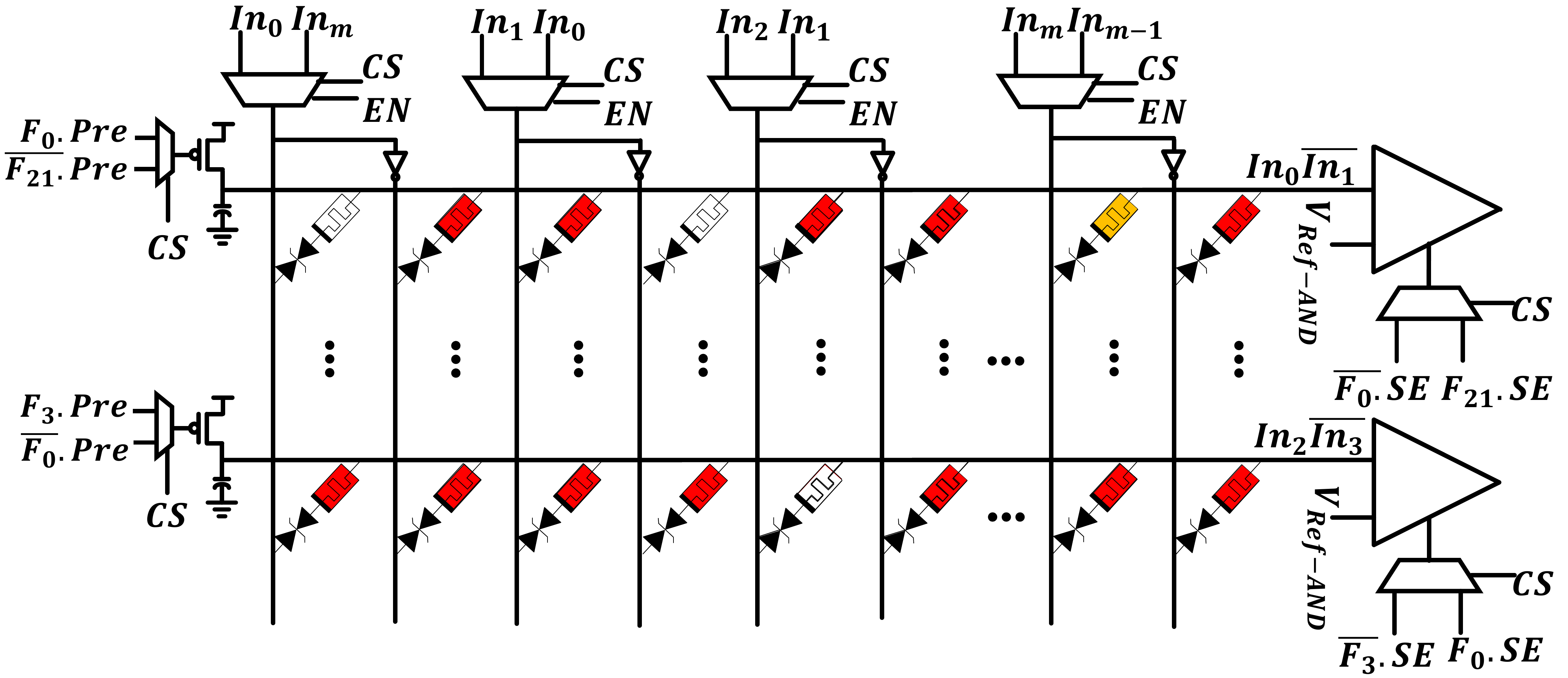}
	\caption{SATO fault mitigation technique.}
	\label{SATO}
\end{figure}

SATO cannot handle faults that appear on two consecutive sets of BLs (each three consecutive BLs are a set if an adder/subtractor is implemented). More multiplexers are needed for each WL to handle faults on consecutive sets of BLs. The number of multiplexers per WL increases linearly with the number of consecutive faulty sets of BLs to be handled by SATO. For example, if faults occur on two consecutive sets of BLs (e.g., if $BL_3$ in Fig. {\ref{SATO}} also contains a fault) SATO cannot handle it unless two or more multiplexers are dedicated to each WL. The probability of two faults occurring on two consecutive BL is less than 3\% for a 64*32 crossbar array. However, SATO is able to handle less than 50\% of the faults if a yield of 99.5\% is considered on a crossbar array. 

\subsubsection{Handling multiple faults}
SATO's efficiency degrades for increasing number of faults. In this paper, we considered a yield of 99.5\% in a 64*32 crossbar array for SATO simulations. This corresponds to 11 randomly distributed faults throughout the array. SATO is able to mitigate around 50\% of the faults in the array. Faults have been distributed on the memory cells using $rand$ function provided by $C++$ programming language.

\begin{figure*}
        \centering
        \begin{subfigure}[b]{0.55\textwidth}
                \centering
                \includegraphics[width=0.99\linewidth]{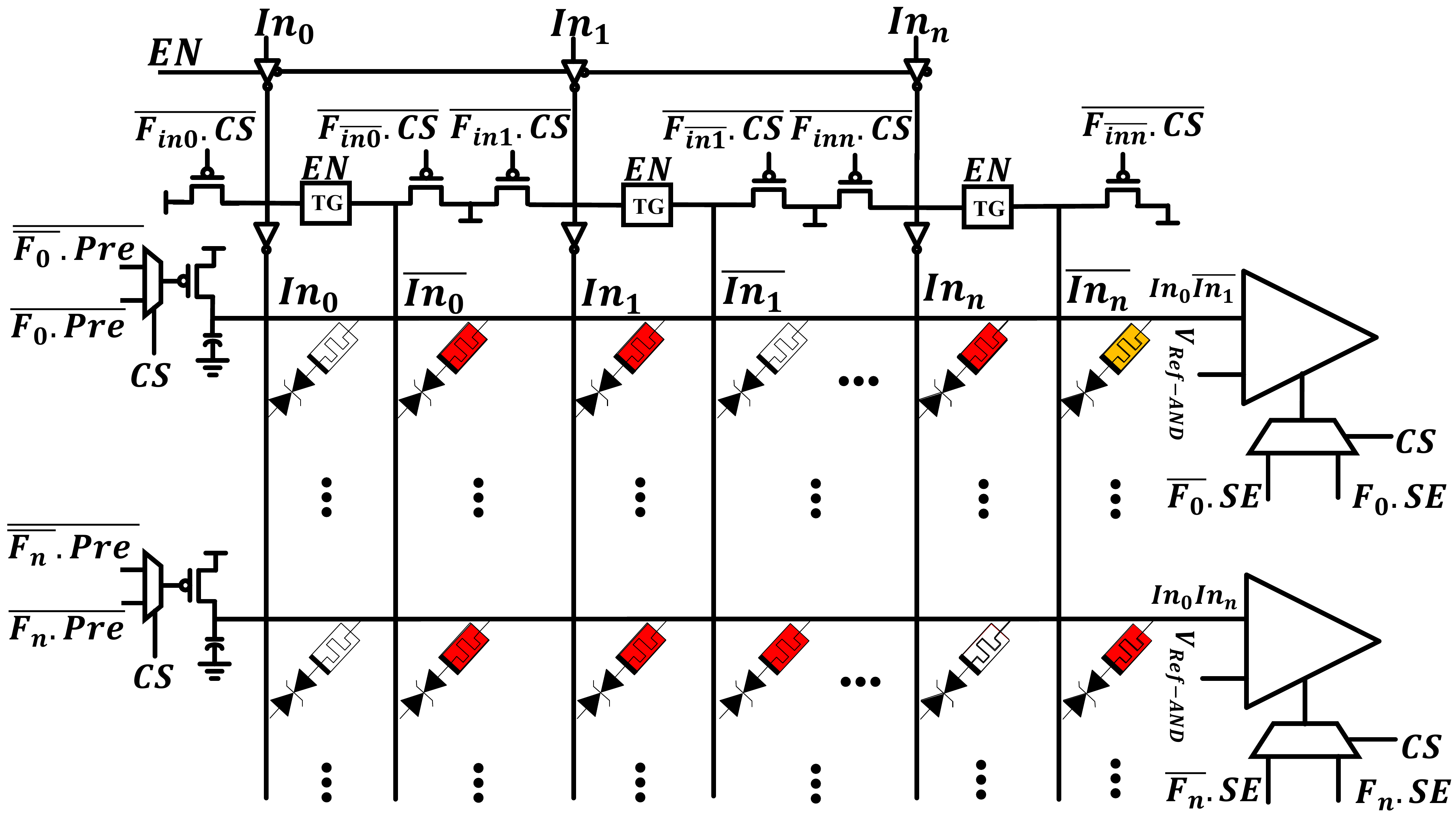}
                \caption{}
                \label{FTV-a}
        \end{subfigure}%
        \begin{subfigure}[b]{0.45\textwidth}
                \centering
                \includegraphics[width=0.99\linewidth]{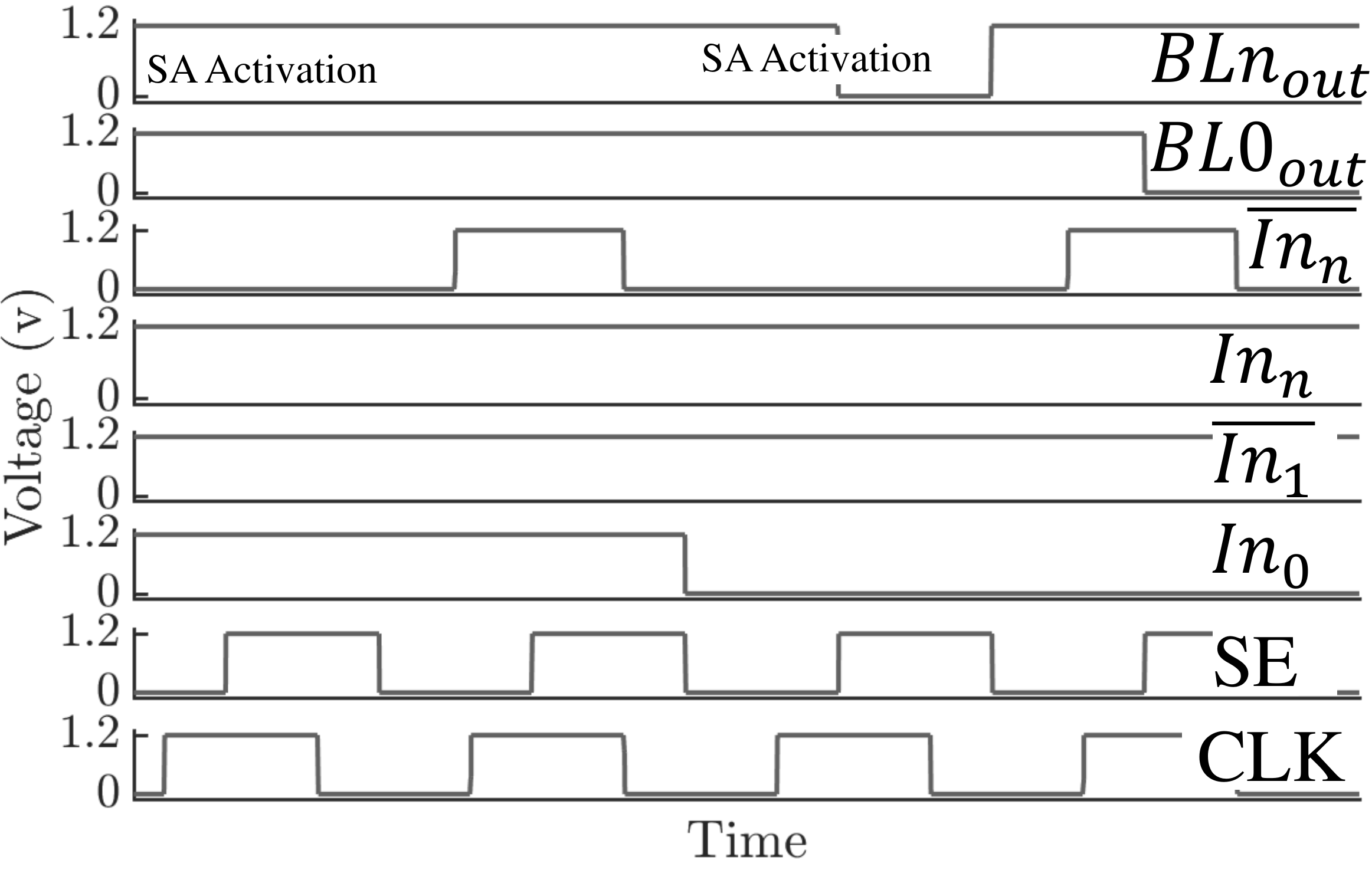}
                \caption{}
                \label{FTV-b}
        \end{subfigure}%
        \caption{FTV: (a) fault mitigation in undesired LRS RRAMs; (b) timing diagram.}
        \label{FTV}
\end{figure*}

\subsection{Forcing to $V_{DD}$ (FTV)}
FTV performs operations of fault-free and faulty BLs in the first and second cycle, respectively. Inputs of faulty RRAMs are forced to $V_{DD}$ in the second cycle. To apply FTV to $NAND$ arrays, we follow a simple $NAND$ logic where for example $A \cdot B \cdot C$ is replaced with $A\cdot B\cdot1$, where C is the input of the SA1 RRAM. Therefore, $NAND-2$ is performed in $NAND-3$ form with an extra `1' which do not affect the logic. However, increased number of RRAMs in a BL reduces the SM. If the faults are located on different BLs, they do not affect the SM.

FTV uses a multiplexer for the enable signal of SAs to ensure that the array is capable of working in two cycles. $\overline{F} \cdot \overline{CS}$ and $F \cdot CS$ are inputs of the multiplexer, where $F$ is `0' if the BL is fault-free and is `1' if the BL is faulty. $CS$ is the clock sequence initialized to `0' in the first cycle and `1' in the second cycle. Enable signal of SAs connected to fault-free BLs are asserted in the first cycle while the enable signal of SAs connected to faulty BLs is asserted in the second cycle to save power and maintain the correct logic.
Furthermore, FTV uses 4 additional transistors compared to DCIM at the WL input to enable the test procedure (explained in Section \ref{test-ftv}) to find faulty RRAMs. 

As shown in Fig. {\ref{FTV-a}}, FTV uses two transmission gates to connect $input$ and $\overline{input}$ to WLs. Also, one PMOS transistor is added to each WL to force the WL to $V_{DD}$ when is needed in the second cycle. In the second cycle, SC signal of faulty WLs gets activated to force faulty WLs to $V_{DD}$. FTV employs a fault signal ($F_W$) for each WL to track the faulty WLs and set them to $V_{DD}$ in the second cycle. FTV also defines a fault signal ($F_B$) for each BL to keep track of faulty BLs. Additional circuitry and peripherals needed to apply FTV to the DCIM increase the area by 9.5\%.


\subsubsection{Forcing-to-Ground (FTG)}
The basic concept of FTG is similar to FTV but it applies to $NOR$ ($OR$) arrays. FTG follows simple logic that number of `0's is not important in $NOR$ ($OR$) operation. Therefore, FTG forces inputs of faulty RRAMs to the ground. Peripherals and the rest of the FTG's operation are the same as FTV.

\subsubsection{Non-fixable Faults }
\begin{figure}
	\centering
	\includegraphics[width= \linewidth]{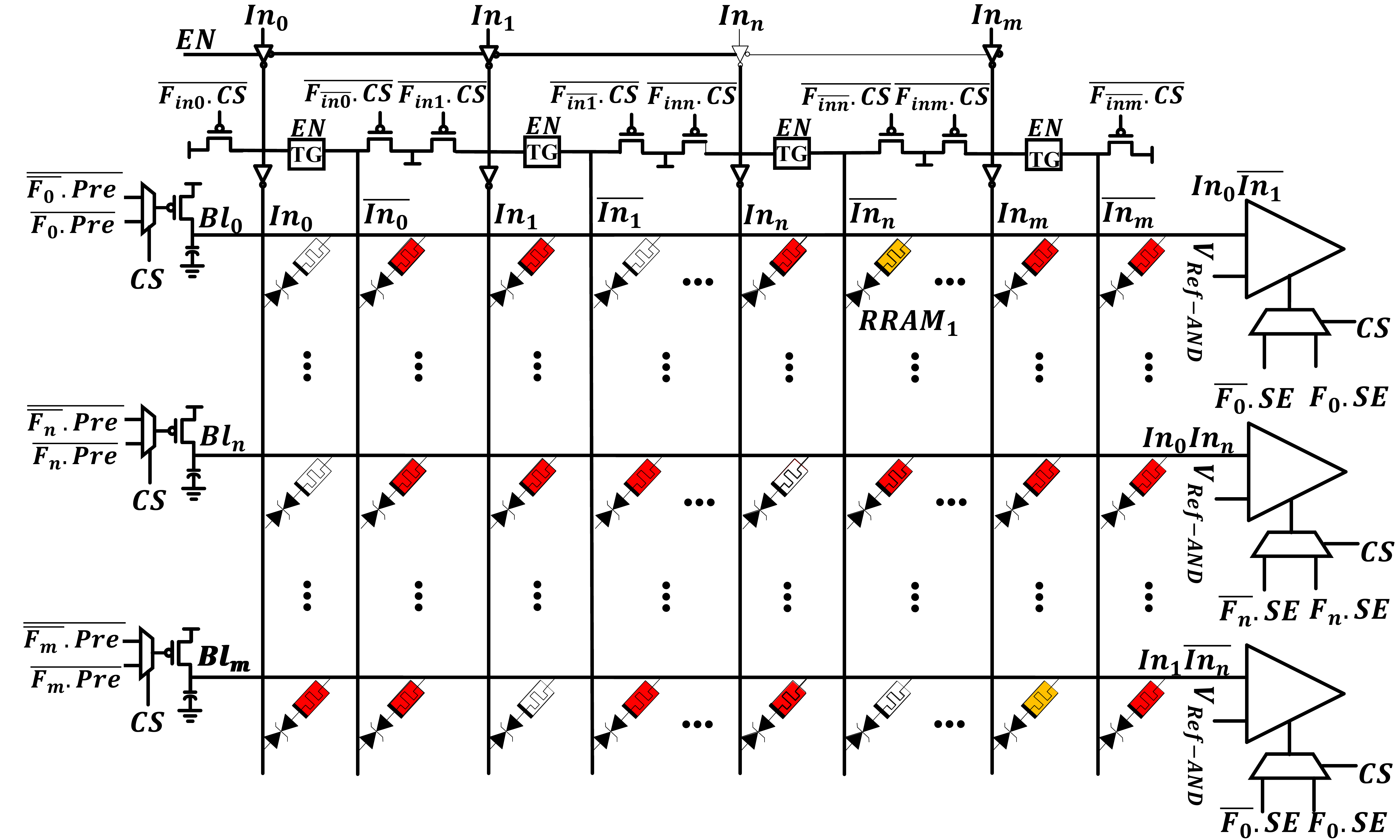}
	\caption{Faults that cannot be handled by FTV.}
	\label{contention}
\end{figure}

Although FTV can fix most of the faults in a crossbar array, it is unable to handle some rare situations. For example, if there are two faulty BLs and the faulty RRAM on one of the BLs is the operand of the other BL. As shown in Fig. {\ref{contention}}, $BL_0$ and $BL_m$ are faulty and their operation must be done in the second cycle. In this case, the logic of $BL_m$ gets lost if FTV forces input of the faulty RRAM (RRAM1) on $BL_0$ to $V_{DD}$ since one of its inputs is set to $V_{DD}$. The $NAND$ operation for $BL_m$ is incorrectly performed between $in_n$ and `1' instead of between $in_n$ and $in_m$. The probability of occurrence of such a fault for fabrication yields of more than 99\% is less than 1\%. We randomly distributed the faults for 100 times using $C++$ language rand function in order to achieve the percentage of faults occurring in an array.

\subsubsection{Handling multiple faults}
As long as faulty RRAMs in the crossbar array are independent of each other, FTV can handle as many as possible faults. For our simulations we inserted 30 faults in a 64*32 crossbar array and FTV was able to solve more than 99\% of fault distribution over the array.

\begin{table}
    \centering
    \caption{Comparison between SATO and FTV }
    \begin{tabular}{|c|c|c|}\hline
        Characteristics & SATO & FTV  \\ \hline
        Coverage & 50\% & 90\% \\ \hline
        SM (w/ diode) & Not Affected & Not Affected \\ \hline
        SM (w/o diode) & Not Affected & Lower \\ \hline
        Test circuitry & Needed & Included \\ \hline
        Area overhead & 28.5\% & 9.5\% \\ \hline
        Power & Not affected & Lower \\ \hline
        Energy & Higher & Slightly Lower \\ \hline
        Performance & 50\% & 50\% \\
        \hline\end{tabular}
        \label{comparison}
\end{table}

\subsection{Finding Faults using FTV Peripherals} {\label{test-ftv}}
It is required to find the faulty RRAMs to set fault signals of the BLs and WLs. Faults can be found by the peripherals that are included in the FTV. However, the BLs must be tested one at a time. To find the faults in a $NAND$ array, input of each RRAM, which is set to LRS in a BL is forced to $V_{DD}$ and the rest are forced to `0'. The output of the SA indicates whether a BL is faulty (`1’) or fault-free (`0’). If the BL is faulty, we need to find out which RRAM is faulty on that BL. A divide-and-conquer approach cannot be used in this architecture since there might be more than one faulty RRAM per BL, so we use brute force algorithm to find faulty RRAM. The input of the RRAM-under-test is set to `0' while inputs of all other RRAMs are set to $V_{DD}$. If SA output is `0', the RRAM-under-test is deemed faulty and its flag is set to `1'. All faulty RRAMs can be found by repeating this operation for each RRAM in each BL sequentially.

\subsection{Usage and Limitations of SATO/FTV/FTG}
SATO/FTV/FTG should be enabled only when a fault has been detected in the test process. Therefore, the fault-free array will only incur area overhead but no performance loss. The faulty array will be salvaged at the cost of performance overhead. 
Note that SATO/FTV/FTG are only applicable to DCIM-based IMC. They cannot be applied to MAGIC or RRAM-based static IMC in the current form. 


\section{Simulation Results} {\label{Fault-Simulation-Results}}
To evaluate SATO and FTV, we compute performance metrics that include worst-case SM,  BL-delay, average delay, average power, and energy consumption of a 64*32 DCIM RRAM crossbar array (\ref{comparison}). Based on the simulation results, FTV is more efficient than SATO.

\begin{figure}
	\centering
	\includegraphics[width= \linewidth]{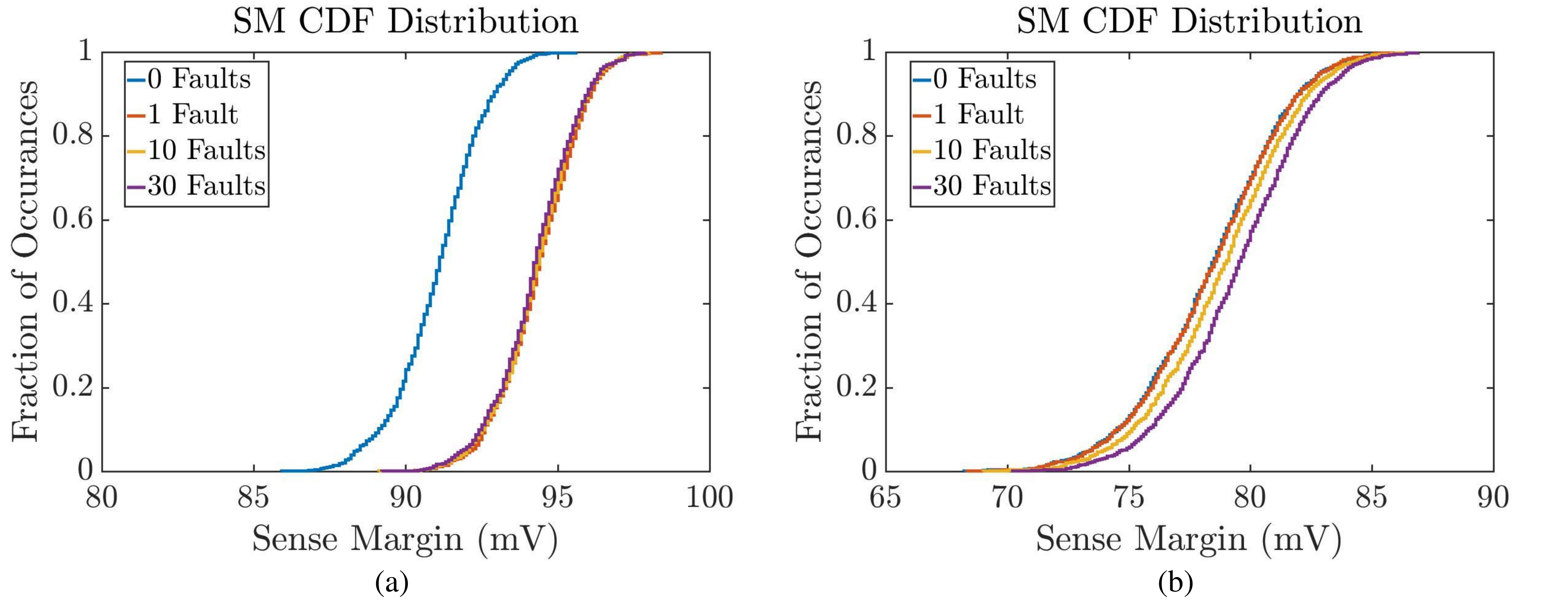}
	\caption{Process variation analysis of SM for various number of failures on a single BL with selector diode in the bitcell i.e., selector diode-RRAM crossbar at, (a) $-10^\degree C$; (b) $90^\degree C$.}
	\label{pvwb}
\end{figure}

\begin{figure}
	\centering
	\includegraphics[width= \linewidth]{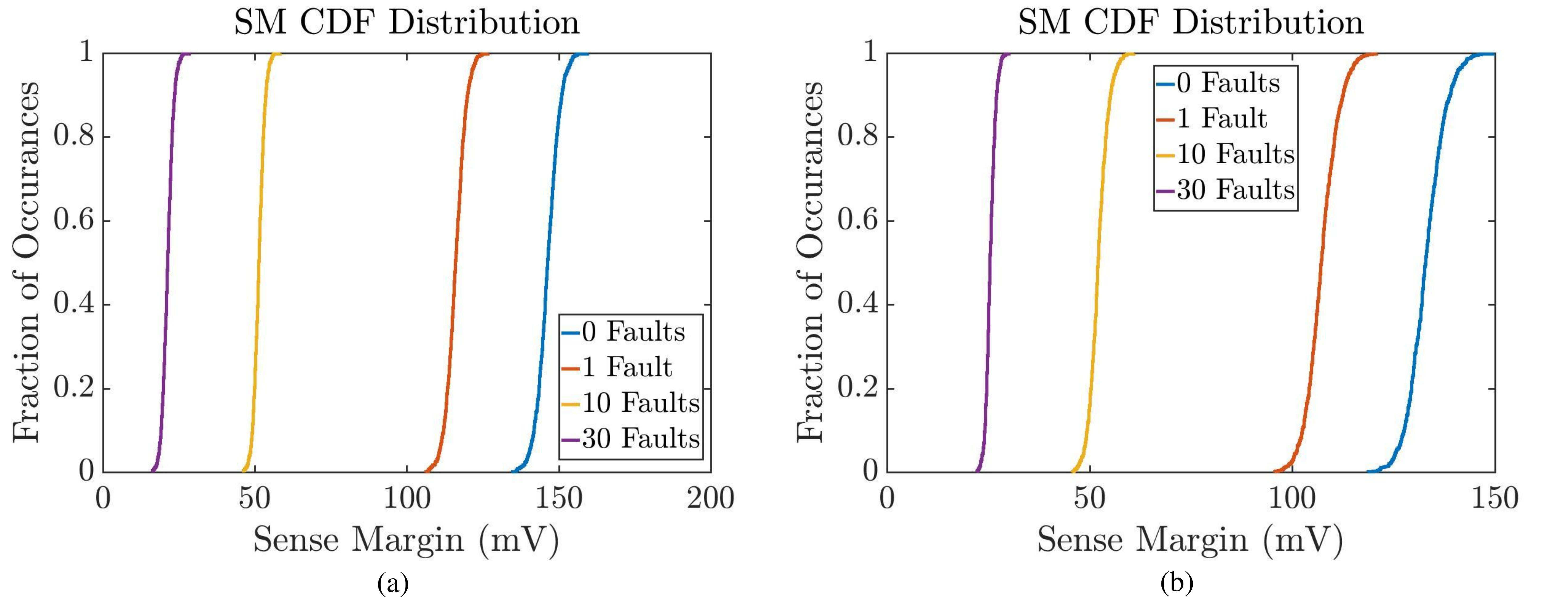}
	\caption{Process variation analysis of SM for different number of failures on a single BL while bitcell consists of RRAM only at, (a) $-10^\degree C$; (b) $90^\degree C$.}
	\label{pvwob}
\end{figure}

\begin{table}
    \centering
    \caption{SATO power and energy consumption}
    \begin{tabular}{|c|c|c|c|c|c|c|c|c|}\hline
        \scriptsize{\# of Faults} & 0 & 1 & 3 & 5 & 10 & 15 & 20 & 30 \\ \hline
        \scriptsize{Power (uW)} & 215.4 & 231.4 & 233 & 234 & 238 & 239 & 240 & 241.5 \\ \hline
        \scriptsize{Energy (pJ)} & 4.3 & 9.4 & 9.4 & 9.5 & 9.6 & 9.7 & 9.7 & 9.8 \\ 
        \hline\end{tabular}
        \label{SATO-parameters}
\end{table}

\subsection{SATO Simulation Results}
Applying SATO to DCIM increases power and energy consumption by 12\% and 127\%, respectively (the worst case) and also performance is reduced by more than 50\%. However, SATO is able to handle $\sim$50\% of the SA1 faults. SAs are very costly and occupy large area, which using SAs to shift data, increases power consumption and leads to higher energy consumption. Power and Energy consumption of SATO with different number of SA1 faults is reported in Table \ref{SATO-parameters}.

\subsection{FTV/FTG Simulation Results}
\begin{figure}
	\centering
	\includegraphics[width= \linewidth]{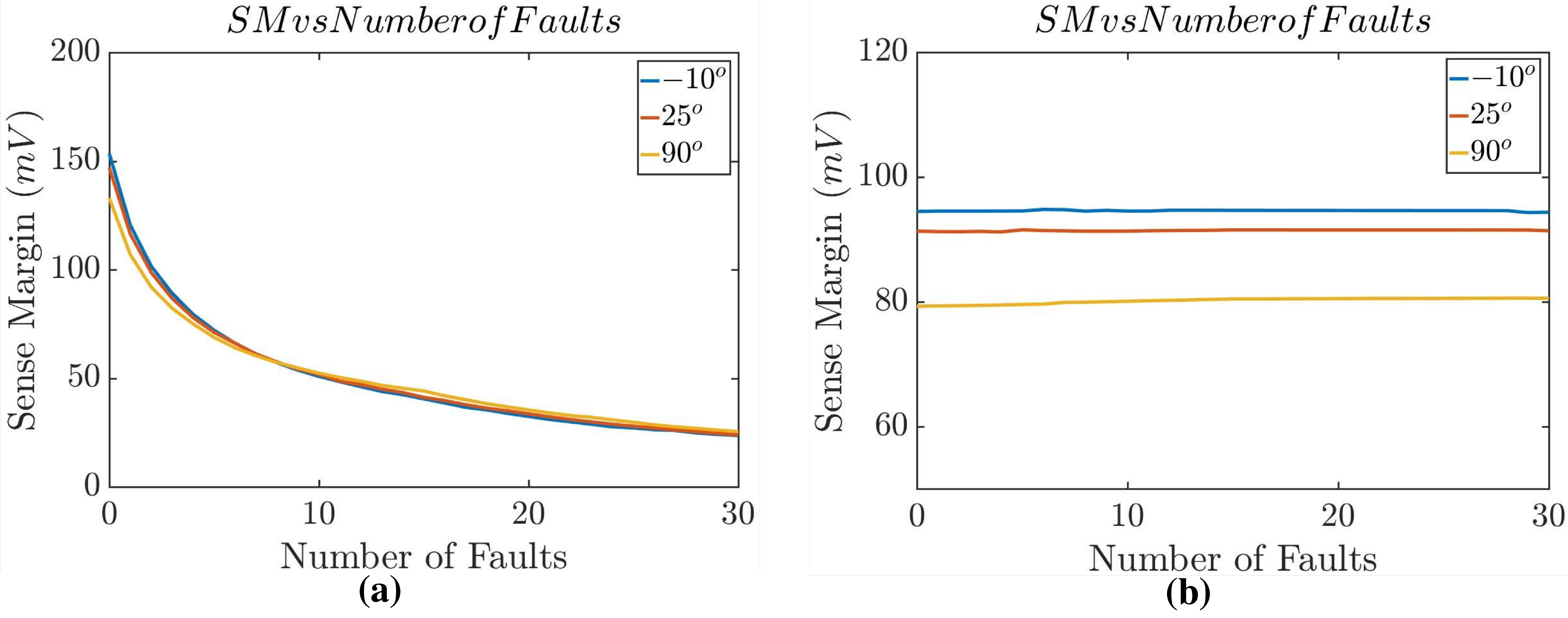}
	\caption{SM for different number of faults in one BL at various temperatures, (a) diode+RRAM in the bitcell; (b) pure RRAM in the bitcell).}
	\label{Temp-figs}
\end{figure}

SM is the most important parameter when FTV is applied to DCIM. Increased number of LRS RRAMs connected to $V_{DD}$ (ground) on a BL worsens the SM when `0' (`1') is the output. 
Considering $NAND-2$, the worst case `0' occurs when one of the operands is `0' and the other operand is `1'. In this case, there is a voltage division is between one LRS RRAM connected to `0' and one LRS RRAM which is connected to $V_{DD}$. When there is a faulty LRS RRAM on the BL and it is forced to $V_{DD}$, the worst case `0' is when two LRS RRAMs are connected to $V_{DD}$ and one LRS RRAM is connected to `0' which lead to increased output `0' voltage on the BL. Increased BL voltage for the worst case `0' degrades the SM as shown in Fig. {\ref{Temp-figs}} (a). 

The degradation in worst case SM happens when the bitcell is made of only a RRAM (i.e., no selector diode).  However, DCIM employs a bidirectional diode in series with the RRAM. This series-connected bidirectional diode is included to reduce power consumption by dropping 0.5V across the 2 terminals. When the voltage difference between a BL and a WL is less then the selector diode threshold voltage there is no current between the BL and the WL. So, increased number of LRS inputs connected to $V_{DD}$ (ground) does not affect the SM of DCIM (Fig. {\ref{Temp-figs}} (b)). 
The current of HRS RRAMs increases with temperature which results in higher sneak path currents. In an $AND$ ($OR$) array, the higher sneak path currents pull up (down) BL voltage to degrade the SM. However, when number of faults increases, sneak paths currents become negligible compared LRS RRAMs which are connected to $V_{DD}$ (ground). Simulation results (Fig. \ref{Temp-figs} (a)) show that the SMs in different temperatures become equal when the number of faults is more than 20. 

Compared to the fault-free situation, FTV reduces power and energy consumption by $>$54\% and $>$7\% respectively (since, SAs consume a lot of power and in the case of FTV, SAs connected to faulty BLs are deactivate in the first cycle and SAs connected to fault-free BLs are deactivated in the second cycle). This is due to inactive BLs and SAs and a longer time of operation. However, the performance reduces by 50\% due to two cycle operations. Average power and energy for the four consecutive $AND$ operations in the 64*64 array are reported in Table {\ref{FTV-parameters}}.

\begin{table}
    \centering
    \caption{FTV power and energy consumption}
    \begin{tabular}{|c|c|c|c|c|c|c|c|c|}\hline
        \scriptsize{\# of Faults} & 0 & 1 & 3 & 5 & 10 & 15 & 20 &30 \\ \hline
        \scriptsize{Power (uW)} & 215.4 & 90 & 91.7 & 92.6 & 96.7 & 97.8 & 99 & 100.1 \\ \hline
        \scriptsize{Energy (pJ)} & 4.3 & 3.6 & 3.7 & 3.7 & 3.9 & 3.9 & 4.0 & 4.0 \\ 
        \hline\end{tabular}
        \label{FTV-parameters}
\end{table}

\subsection{Process Variation Simulations}
The most important parameter to consider in a crossbar array under process variation is SM. We ran 1000-point MC simulations at $-10^\degree C$, and $90^\degree C$ on DCIM by considering the bitcell consisting of only a RRAM and RRAM and a selector diode with different number of SA1 faults. Simulation results for RRAM bitcell and RRAM and selector diode bitcell are shown in Fig. {\ref{pvwb}} and {\ref{pvwob}}, respectively.   

As shown in Fig. \ref{pvwb}, variations do not affect SM significantly due to the presence of selector diode which stabilizes BL voltage. However, as demonstrated in Fig. \ref{pvwob}, variations affect the SM when only RRAM is used in the crossbar. This is due to large changes in the RRAM resistance for a small change in RRAM gap when 1.2V is applied across it.
Worst case SM with the number of failures for both w/ and w/o selector diode is reported in Table. \ref{sm-pv}.

\begin{table}
    \centering
    \caption{SM for different number of failures}
    \begin{tabular}{|c|c|c|}\hline
        Failures & SM (Selector diode) & SM (Without Selector Diode) \\ \hline
        0 &  91.3 mV & 118 mV \\ \hline
        1 &  91.2 mV & 95 mV \\ \hline
        10 & 91.3 mV & 44 mV  \\ \hline
        30 & 91.4 mV & 19 mV  \\ 
        \hline\end{tabular}
        \label{sm-pv}
\end{table}

\section{Conclusions} {\label{Conclusion}}
We proposed FAME for in-memory FP arithmetic computation. FAME implements single precision FP adder/subtractor using RRAM crossbar and evaluated two flavors with $NAND-NAND$ and $NOR-NOR$ compute arrays. We also proposed a novel SA based shift circuit for frequent shifting needed in FP operation. Compared to MAGIC-based implementation, FAME achieves 828X and 3.7X latency and energy improvement over MAGIC and compared to processing units (e.g. CPU, FPGA, GPU) it also reduces energy consumption and delay by $93\%$ and $70\%$, respectively.
FAME achieves lower power and energy consumption compared to MAGIC and processing units at low area overhead to the memory arrays. FAME uses 3KB memory to implement single precision FP operations ({\ref{FAME-Area}}). 
Furthermore, two approaches to mitigate HRS to LRS retention and stuck-at-1 failures in RRAM-based compute memories are proposed along with a test approach to identify faulty RRAMs. Forcing-to-$V_{DD}$ (FTV) can mitigate 99\% of the faults while reducing the power consumption by $>$50\% and energy consumption by $>$7\%. Shifting-at-the-Output (SATO) technique increases power consumption slightly but increases energy consumption by $>$50\%. 

\vspace{2 mm}
\noindent \textbf{Acknowledgement:} This work is supported by SRC (2847.001), and NSF (CNS- 1722557, CCF-1718474, CNS-1814710, DGE-1723687 and DGE-1821766).

\bibliographystyle{IEEEtran}
\bibliography{IEEEabrv,FPIMC}
\end{document}